\documentstyle[aps,prl,multicol,epsf]{revtex}
\begin{document} 

\title{Field Theory of Mesoscopic Fluctuations in
  Superconductor/Normal-Metal Systems}

\author{Alexander Altland, B D Simons, and D Taras-Semchuk} 
\address{Cavendish Laboratory, Madingley Road, Cambridge CB3\ OHE, UK}

%\date{\today}
\maketitle 
 
\begin{abstract} 
  Thermodynamic and transport properties of normal disordered
  conductors are strongly influenced by the proximity of a
  superconductor. A cooperation between mesoscopic coherence and
  Andreev scattering of particles from the superconductor generates
  new types of interference phenomena. We introduce a field theoretic
  approach capable of exploring both averaged properties {\it and
    mesoscopic fluctuations} of superconductor/normal-metal systems.
  As an example the method is applied to the study of the level
  statistics of a SNS-junction.
\end{abstract} 

%\pacs{PACS numbers: }

\begin{multicols}{2}  
%
%\end
Physical properties of both superconductors and mesoscopic normal
metals are governed by mechanisms of macroscopic quantum coherence.
Their interplay in hybrid systems comprised of a superconductor
adjacent to a normal metal gives rise to qualitatively new phenomena
(see Ref.~\cite{Beenakker} for a review): Aspects of the superconducting
characteristics are imparted on the behaviour of electrons in the
normal region. This phenomenon, known as the ``proximity effect'',
manifests itself in a) the {\it mean} (disorder averaged) properties of
SN-systems being substantially different from those of normal metals
and b) various types of mesoscopic {\it fluctuations} which not only
tend to be larger than in the pure N-case but also can be of
qualitatively different physical origin.  Although powerful
quasi-classical methods, based largely on the pioneering
work of Eilenberger~\cite{Eilenberger} and Usadel~\cite{Usadel},
have been developed to analyse the
manifestations of the proximity effect in average characteristics of
SN-systems, far less is known about the
physics of mesoscopic fluctuations: While the quasi-classical
approach is not tailored to an analysis of fluctuations, standard
diagrammatic techniques~\cite{Altshuler} used in the study of
N-mesoscopic fluctuations can often {\it not} be applied due to the
essentially non-perturbative influence of the fully established
proximity effect. Important progress was made recently by extending
the scattering formulation of transport in N-mesoscopic systems to the
SN-case~\cite{Beenakker}. This approach has proven powerful in the
study of various transport fluctuation phenomena but is not applicable
to the study of fluctuations on a local and truly microscopic level.

In the present Letter we introduce a general framework that combines
key elements of the quasi-classical approach with more recent methods
developed in N-mesoscopic physics into a unified approach. As a result
we obtain a formalism that can be applied to the general analysis of
mesoscopic fluctuations superimposed on a proximity effect influenced
mean background. In order to demonstrate the practical use of the
approach we will consider the example of {\it spectral fluctuations}
as a typical representative of a mesoscopic phenomenon. The density of
states (DoS) of N-mesoscopic systems exhibits quantum fluctuations
around its disorder averaged mean value which can be described in
terms of various types of universal statistics. The analogous question
in the SN-case -- What types of statistics govern the disorder induced
fluctuation behaviour of the {\it proximity effect influenced} DoS? --
has not been answered so far. Although space limitations prevent us
from discussing this problem in detail, our main result, the emergence
of some kind of modified Wigner Dyson statistics\cite{Mehta}, will be
derived below.

%***************************************
%* Figure:  SNS junction               *
%***************************************
\narrowtext
\begin{figure}[hbt]
\centerline{\epsfxsize=2.5in\epsfbox{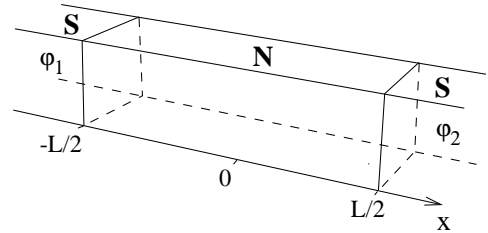}}
\caption{Geometry of the SNS-junction.}
\label{fig:sns}
\end{figure}

To be specific we consider the geometry shown in Fig.~\ref{fig:sns},
where $\varphi=\varphi_1-\varphi_2$ represents the relative phase of
the complex order parameters of the two superconductors. It is well
known~\cite{review} that even the mean DoS of the SNS-system exhibits
non-trivial behavior which is difficult to describe within standard
perturbative schemes: States which fall within the superconducting
gap, $\Delta$, are confined to the normal metal. The proximity effect
then further induces a minigap in the DoS of the {\it normal} region
around the Fermi-energy, $\epsilon_F$, whose size of $O(E_c=D_n/L^2)$
depends sensitively on $\varphi$ ($D_n$ is the diffusion constant and
$\hbar=1$ throughout).  To analyse the fluctuation behavior of the
DoS, $\nu(\epsilon)$, around its disorder averaged background, $\langle
\nu(\epsilon)\rangle$, we will consider the two-point correlation
function~\cite{fn1},
\begin{eqnarray}
R_2(\epsilon,\omega)=\left\langle\nu(\epsilon)\right\rangle^{-2}
\left\langle \nu(\epsilon+\omega/2)\nu(\epsilon-\omega/2)\right\rangle_c.
\label{R2}
\end{eqnarray}
The starting point of our analysis is the Gor'kov equation for the
matrix advanced/retarded (a/r) Green function~\cite{Eilenberger}
\begin{eqnarray}
{\cal G}^{r,a}_\epsilon=\left(\matrix{G^{r,a}_\epsilon&F^{r,a}_\epsilon\cr 
{F^\dagger}^{r,a}_\epsilon&{G^\dagger}^{r,a}_\epsilon}\right),
\label{gorkov}
\end{eqnarray}
where
\end{multicols}
\widetext
\begin{equation}
\left[\epsilon_F-{1\over 2m}\left(\hat{p}-{e\over c}
\vec{A}\sigma_3^{\rm ph}\right)^2-
V(\vec{r})+\left(\hat{\Delta}(\vec{r}) +\epsilon_\pm\right)\sigma_3^{\rm ph}
\right]{\cal G}^{r,a}_\epsilon(\vec{r},\vec{r'})=\delta^d(\vec{r}-\vec{r'}),
\label{schroedinger}
\end{equation}
\vspace{-0.5cm}
\begin{multicols}{2}
\narrowtext\noindent
$\epsilon_\pm\equiv\epsilon\pm i0$, $\vec{A}$ is the vector
potential of an external magnetic field, $\hat{\Delta}=\Delta
\sigma_1^{\rm ph} \exp(-i\varphi \sigma_3^{\rm ph})$ represents the
(spatially dependent) complex order parameter with phase $\varphi$,
and Pauli matrices $\vec{\sigma}^{\rm ph}$ operate in the Nambu or
particle/hole (ph) space. The impurity potential in the N-region is
taken to be Gaussian $\delta$-correlated with zero mean and
correlation $\left\langle V(\vec{r}) V(\vec{r}') \right \rangle =
\delta^d \left( \vec{r}-\vec{r}'\right)/(2 \pi\nu\tau)$, where $\nu$
denotes the DoS of the bulk normal metal at $\epsilon_F$, and $\tau$
represents the mean free scattering time. In the following the complex
order parameter in the S-region is {\it imposed} and not obtained
self-consistently\cite{fn6}. Where the S- and N- region are distinct
(as in the SNS junction), the bulk DoS, $\nu_{n,s}$ and scattering time,
$\tau_{n,s}$ will be chosen independently.

Traditionally the {\it impurity averaged} Green
function~(\ref{gorkov}) is computed within a quasi-classical
approximation, i.e. the Schr{\"o}dinger equation~(\ref{schroedinger})
is reduced to an effective transport equation, the Eilenberger
equation~\cite{Eilenberger}, which in the dirty limit simplifies
further to the diffusive Usadel equation~\cite{Usadel}.  Here we
develop a field theoretical formulation that integrates concepts of
the quasi-classical formalism into a more general framework allowing
for the computation of disorder averaged {\it products of} Green
functions, a necessary requirement for the calculation of correlation
functions such as~(\ref{R2}). The basic strategy will be to start from
a (microscopically derived) generating functional whose points of
stationary phase obey the Usadel equation. By investigating
fluctuations around this quasi-classical limit, correlations between
the different Green functions will be explored.  In the following we
formulate this program in more detail.

As in the pure N-case, ensemble averaged products of advanced and
retarded Gorkov Green functions can be described in terms of
generating functionals of nonlinear $\sigma$-model
type\cite{Oppermann} (see Ref.\cite{Efetov83} for a review on the
$\sigma$-model analysis of Green functions in N-mesoscopic physics).
In the dirty limit, $(\epsilon,\Delta)<\tau^{-1}\ll \epsilon_F$, the
generalization of the supersymmetric N-type
$\sigma$-model\cite{Efetov83} reads
\begin{eqnarray}
\label{funint}
&&\hspace{2.5cm}\int_{Q^2=\openone} DQ (\cdots) e^{-S[Q]},\\ 
&&S[Q]=-{\pi\nu\over 8}\int {\rm str}\Big[D(\widetilde{\partial}Q)^2
+4iQ\big(\widetilde{\Delta}+\epsilon+{\omega_+\over 2}\sigma_3^{\rm ar}
\big)\sigma_3^{\rm ph}\Big],\nonumber
\end{eqnarray}
where $\widetilde{\partial}=\partial-i(e/c)[\vec{A_\phi}\sigma_3^{\rm
  tr} \otimes \sigma_3^{\rm ph}, \cdot]$ represents a covariant
derivative, $\vec{A_\phi} = \vec{A} +c/(2e) \partial \phi$ accounts
for both the external field and the phase of the order parameter,
$\widetilde{\Delta}=\Delta \sigma_2^{\rm ph}$, the Pauli matrices
$\vec{\sigma}^{\rm fb}$, $\vec{\sigma}^{\rm tr}$ and
$\vec{\sigma}^{\rm ar}$ operate in fermion/boson, time-reversal and
ar-blocks respectively\cite{Efetov83}. The symbol $D$ stands for a
space dependent diffusion constant which may take separate values,
denoted as $D_{n,s}$, in the N and S regions.  Although specific
pre-exponential source terms (denoted by ellipses in
Eq.(\ref{funint})) must be chosen according to any given correlation
function (such as (\ref{R2})), their precise form does not influence
the analysis below and we therefore refer to Ref.~\cite{Efetov83} for
their detailed structure.  The integration in (\ref{funint}) extends
over a $16 \times 16$-dimensional matrix field $Q=T^{-1} \sigma_3^{\rm
  ph} \otimes \sigma_3^{\rm ar} T$, whose symmetries are identical
with those of the conventional $\sigma$-model\cite{Efetov83}.

The expression~(\ref{funint}) differs in two respects from the
$\sigma$-model for N-systems: i) the appearance of a ph-space
associated with the $2\times2$-matrix structure of the Gorkov Green
function, and ii) the presence of the order parameter $\tilde{\Delta}$.
Whereas i) can be accounted for by a doubling of the matrix dimension
of the field $Q$, ii) calls for more substantial modifications: For
$\Delta\not=0$ standard perturbative schemes for the evaluation of the
functional (\ref{funint}) fail\cite{fn2}, an indication of the fact
that the superconductor influences the properties of the normal metal
heavily. Under these conditions a more efficient approach is first to
subject the action to a mean field analysis and then to consider
fluctuations around a newly defined --- and generally space dependent
--- stationary field configuration. A variation of the action
(\ref{funint}) with respect to $Q$, subject to the constraint $Q^2=
\openone$, generates a non-linear equation for the saddle-point,
\begin{eqnarray}
D\widetilde{\partial}_i(\bar{Q}\widetilde{\partial}_i \bar{Q})+
\left[\bar{Q},
  \Delta \sigma_1^{\rm ph} -i(\epsilon + \omega_+\sigma_3^{\rm
    ar})\sigma_3^{\rm ph}\right]=0. 
\label{Usad}
\end{eqnarray}
Current conservation implies the boundary condition~\cite{Kuprianov},
$\sigma_n Q\partial_x|_{x_n} Q=\sigma_s Q \partial_x|_{x_s} Q$, where
$\sigma_{n,s}=e^2\nu_{n,s} D_{n,s}$ denotes the conductivity and
$\partial_x|_{x_{n(s)}}$ is a normal derivative at
the N(S)-side of the interface.

An inspection of~(\ref{Usad}) shows that only the particle/hole
components of the matrix field $\bar Q$ are coupled by the saddle
point equation. It is thus sensible to make a block diagonal ansatz
$\bar Q={\rm bdiag\,}(q_+,q_-)$, where the eight dimensional retarded,
$q_+$ and advanced, $q_-$ subblocks are diagonal in both time reversal
and boson/fermion space.  Noting that the saddle point configuration
$-i\pi\nu q_\pm$ of the nonlinear $\sigma$-model is associated with the
impurity averaged retarded/advanced Green function~\cite{Efetov83}, we
identify Eq.~(\ref{Usad}) as the Usadel equation.  The general
connection between the $\sigma$-model formalism and the
quasi-classical approach has first been noticed
in~\cite{Muzykantskii95}.

Eq.~(\ref{Usad}), in its interpretation as the Usadel equation, has
been discussed at length in the
literature~\cite{review,Bruder,Spivak}.  Although complex in general,
the solutions have a simple qualitative geometric interpretation:
Employing the explicit parametrization $q_\pm = \vec{q_\pm}\cdot
\vec{\sigma}^{\rm ph}$, Eq.~(\ref{Usad}) describes the gradual rotation of
the three dimensional vector $\vec{q}$ from a direction almost
parallel to $\hat{e}_1$ in the bulk superconductor to a
value aligned with $\hat{e}_3$ deep in the normal metal. Closed
analytical solutions have been obtained for various types of
geometries, including the SNS-structure above.

So far our analysis has been for SN-systems of a general geometry.
Specializing the discussion to the particular SNS-junction shown in
Fig.~(\ref{fig:sns}), we set $\Delta(\vec{r})\equiv\Delta
\Theta(|x|-L/2)$ constant inside the superconductor ($\Delta \gg
E_c$), and zero in the normal region, with a phase $\pi/2+{\rm
  sgn}(x)\varphi/2$. The saddle-point equation depends sensitively on
both the presence or absence of an external magnetic field and the
phase difference between the order parameters.  Taking the external
field to be zero, it is convenient to focus on two extreme cases: (i)
$\varphi=0$ (orthogonal symmetry), and (ii) $ \varphi\gg 1/\sqrt{g}$
(unitary symmetry).  Here $g=E_c/\bar{d}\gg 1$ denotes the
dimensionless conductance and $\bar{d}$ represents the bulk
single-particle level spacing of the normal metal.

\begin{figure}[hbt]
\centerline{\epsfxsize=3.0in\epsfbox{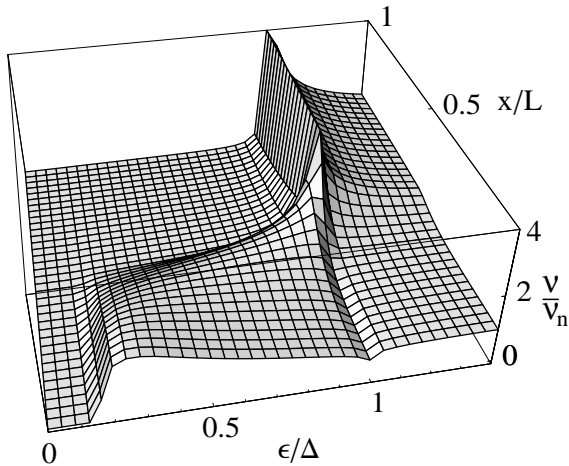}}
\vskip0.4truein
\caption{Local DoS $\nu$ shown as a function of energy and position for 
$\gamma=0.1$, $L/\xi=5$ and $\varphi=0$. }
\label{fig:dos}
\end{figure}

The {\it disorder averaged local DoS} can be obtained from the
analytical solution of the Usadel-saddle point
equation~\cite{review,Bruder,Spivak,fn5,Altland97} as
$\nu(\vec{r})=\nu {\rm Re} [q_+(\vec{r})]_3$. Its space/energy
dependence is shown in Fig.~\ref{fig:dos} for the case (i), zero phase
difference, and a particular value of the material parameter
$\gamma=\nu_n \sqrt{D_n} / \nu_s \sqrt{D_s}$. The most striking
feature of the average DoS is the appearance of a spatially constant
minigap in the N-region.  The gap attains its maximum width $E_c$ at
$\varphi=0$ and shrinks to $0$ as $\varphi$ approaches
$\pi$~\cite{Spivak}.  Within the superconductor, $\nu(E_c < \epsilon
\ll \Delta)$ decays exponentially on a scale set by the bulk coherence
length $\xi=(D_s/2 \Delta)^{1/2}$.

%***************************************
%* Figure:  SNS spectrum               *
%***************************************
\narrowtext

We next turn to the main subject of this Letter, the issue of {\it
  fluctuations around the Usadel saddle-point}. Employing the
parametrization $Q=T^{-1} \bar{Q} T, \; T\not= \openone$, three
qualitatively different types of fluctuations can be identified: (a)
fields $T$ which are diagonal in the space of advanced and retarded
components, (b) $T$'s which commute with all matrices $\sigma_i^{\rm
  ph}$ but mix advanced and retarded components, and (c) $T$'s
fulfilling neither of the conditions (a) and (b). Fluctuations of
(a)-type preserve the ar-diagonal structure of the saddle point. These
fluctuations do not give rise to correlations between advanced and
retarded Green functions. Nonetheless, they are of physical
significance: Quantum corrections to the Usadel solution, most
importantly the renormalization of the minigap by weak localization
effects and the existence of rare prelocalized
states~\cite{Muzykantskii95,Prelocal} below the gap, are described by
fluctuations of this type.  For brevity, we postpone further
discussion of these results to a separate paper\cite{Altland97} and,
instead, turn to the discussion of the second type of fluctuations, (b).

Consider the saddle point equation~(\ref{Usad}) in the simple case
$\omega=\varphi=0$. Obviously, as it commutes with all matrices 
$\sigma_i^{\rm ph}$, any spatially constant rotation $T$ of
type (b) gives
rise to another solution. In other words, the (b)-fluctuations represent
Goldstone modes with an action that vanishes in the limit of spatial
constancy and $\omega\to0$. Since any $T$ diagonal in
ph-space inevitably has to couple between advanced and retarded
indices\cite{Efetov83}, these modes lead to {\it correlations between
  advanced and retarded Green functions} (and thereby to mesoscopic
fluctuations) which become progressively more pronounced as $\omega$
approaches zero.

In the limit of small frequencies $\omega<E_c$, the {\it ergodic
  regime}, the global zero mode $Q_0=T_0^{-1}qT_0$,
$[T_0,\vec{\sigma}^{\rm ph}] = 0, \; T_0( \vec{r}) = {\rm const.}$,
plays a unique role: Whereas fluctuations with non-vanishing spatial
dependence give rise to contributions to the action of
$O(g \gg 1)$\cite{Efetov83}, this mode couples {\it only} to the frequency
difference $\omega$.  Restricting attention to the pure zero mode
contribution, we obtain the effective action
\begin{eqnarray}
S_0[Q_0]=-i{\pi\over 2}{\omega_+\over \bar{d}(\epsilon)} {\rm str}\left[Q_0
\sigma_3^{\rm ar}\right], 
\label{Szero}
\end{eqnarray}
where $\bar{d}(\epsilon)=(\int \nu(\epsilon))^{-1}$ denotes the
average level spacing and the ph-degrees of freedom have been traced
out. From this result it follows\cite{Efetov83} that, in the ergodic
regime, the spectral statistics of an SNS system is governed by
Wigner-Dyson fluctuations~\cite{Mehta,fn3} of (i) orthogonal or (ii)
unitary symmetry superimposed upon an energetically non-uniform mean
DoS.  Furthermore, a comparison of Eq.~(\ref{Szero}) with the
analogous action for N-systems~\cite{Efetov83} shows the correlations
to depend on an average level spacing that is effectively {\em
  halved}. This reflects the strong ``hybridization'' of levels at
energies $\sim \epsilon_F \pm \epsilon$ induced by Andreev scattering
at the SN-interface.  

In further contrast to N-systems, the range over which Wigner-Dyson
statistics apply turns out to be greatly diminished by non-universal
fluctuations. To prove the last statement, fluctuations of type (c),
coupling between advanced/retarded {\it and} particle/hole components
simultaneously, have to be taken into account.  Due to their
complexity, the detailed analysis of the (c)-type fluctuations is
cumbersome and will be deferred to a forthcoming
publication~\cite{Altland97}. Here we restrict ourselves to a brief and
qualitative discussion of the principal effect of these fluctuations
on the universal zero mode action~(\ref{Szero}).

As can be seen from the general structure of the
action~(\ref{funint}), (c)-type fluctuations in the vicinity of the
minigap are generally 'massive' (governed by an action which is at
least of order $\epsilon/\bar{d} \gtrsim g\gg1$). It is thus
permissible to treat these fluctuations in a simple Gaussian
approximation. As a result we arrive at an effective action
which is non-perturbative in the above zero mode configuration and
quadratic in the (c)-type perturbations. Integrating over these fields
in a spirit similar to the analysis performed in Ref.~\cite{Kravtsov}
we obtain the modified zero mode action
\begin{eqnarray}
S[Q_0]=S_0[Q_0]-{\kappa(\epsilon) \over g} \left({\omega\over 
\bar{d}(\epsilon)}\right)^2 {\rm str}\left[\sigma_3^{\rm ar},Q_0\right]^2,
\label{seff}
\end{eqnarray}
where $\kappa(\epsilon)\sim O(1)$ denotes a constant dependent on the
sample geometry\cite{Altland97}. Eq.~(\ref{seff}) has a structure
equivalent to that found in the study of universal parametric
correlation functions and explicit expressions for $R_2$ for both
orthogonal and unitary ensembles can be deduced from
Ref.~\cite{Simons}. Qualitatively, the additional contribution in
(\ref{seff}) counteracts the zero-mode fluctuations for non-vanishing
frequencies $\omega$. Already for energy separations
$\omega/\bar{d}(\epsilon)\sim \sqrt{g}$, the zero-mode integration is
largely suppressed which manifests in an exponential vanishing of the
level correlations on these scales. This is in contrast to the pure
N-case where the Wigner-Dyson regime (prevailing up to frequencies
$\omega\simeq E_c$) is succeeded by other forms of {\it algebraically}
decaying spectral statistics in the high frequency domain
$\omega>E_c$\cite{AS}.

In conclusion a general framework has been developed in which the
interplay of mesoscopic quantum coherence phenomena and the proximity
effect can be explored. An investigation of the spectral statistics of
an SNS geometry revealed that level correlations are Wigner-Dyson
distributed with strong non-universal corrections at large energy
scales. Finally, we remark that for quantum structures in which
transport is not diffusive but ballistic and boundary scattering is
irregular, a ballistic $\sigma$-model involving the classical Poisson
bracket can be derived~\cite{Andreev96}. In this case, the
saddle-point condition recovers the Eilenberger equation of
transport~\cite{Eilenberger}.

We are indebted to Boris Altshuler, Anton Andreev, Konstantin Efetov,
Dima Khmel'nitskii, Valodya Falko and Martin Zirnbauer for useful
discussions. One of us (DT-S) acknowledges the financial support of
the EPSRC. The hospitality of the ITP in Santa Barbara and the Lorentz
Center in Leiden are gratefully acknowledged. This research was
supported in part by the National Science Foundation under Grant No.
PHY94-07194.  

\end{multicols}


\begin{references}
\bibitem{fn6} In general, the proximity of a N-metal leads to a
  renormalization of the superconducting order parameter in the
  boundary region. However, we do not expect this phenomenon to have
  an essential impact on our results.
\bibitem{fn5} Note that the solution of Eq.~(\protect\ref{Usad}) does not
  in general belong to the standard saddle-point manifold of the
  $\sigma$-model. It can however be accessed by a deformation of the
  integration contour\cite{Altland97}.
\bibitem{fn3} With reference to the specific correlation function $R_2$, 
we remark that only massive fluctuations in the ph-sector contribute
to connected correlators of the form $\left\langle {\cal
G}^A_{\epsilon_1} {\cal G}^A_{\epsilon_2}\right\rangle$ allowing such
terms to be neglected.
\end{references}
\end{document}